# Plasmonic Titanium Nitride via Atomic Layer Deposition: A Low-Temperature Route

Dhruv Fomra[1], Ray Secondo[1], Kai Ding[1], Vitaliy Avrutin[1], Natalia Izyumskaya[1], Ümit Özgür[1], Nathaniel Kinsey[1]

Department of Electrical and Computer Engineering, Virginia Commonwealth University, Richmond, Virginia – 23220

**Abstract**

To integrate plasmonic devices into industry, it is essential to develop scalable and CMOS compatible plasmonic materials. In this work, we report high plasmonic quality titanium nitride (TiN) on c-plane sapphire by plasma enhanced atomic layer deposition (PE-ALD). TiN with low losses and high metallicity was achieved at temperatures below 500°C, by exploring the effects of chemisorption time, substrate temperature and plasma exposure time on material properties. Reduction in chemisorption time mitigates premature precursor decomposition at $T_S > 375°C$, and a trade-off between reduced impurity concentration and structural degradation caused by plasma bombardment is achieved for 25s plasma exposure. 85 nm thick TiN films grown at a substrate temperature of 450°C, compatible with CMOS processes, with 0.5s chemisorption time and 25s plasma exposure exhibited a high plasmonic figure of merit ($|\varepsilon'/\varepsilon''|$) of 2.8 and resistivity of 31 $\mu\Omega - cm$. These TiN thin films fabricated with subwavelength apertures were shown to exhibit extraordinary transmission.

Introduction:

Advancements in materials has long been an enabler of next-generation technologies. Gallium nitride in light emitting diodes (LEDs),[1,2] high-k dielectrics in integrated circuits, and high-purity silica leading to optical fibers,[3,4] are just few examples of the major advancements in materials research that have paved the way for breakthrough technologies. Optics is no different, where silicon has given rise to compact on-chip photonic platforms.[5] However, the large disparity in size scales between photons and electrons remains. More recently, plasmonics has been seeking to address this by the coupling of electromagnetic waves to the motion of free electrons on the surface of a metal.[6,7] This enables confinement and control of light at scales much smaller than the diffraction limit by enhancing light matter interaction. A variety of applications including chemical sensors,[8,9] nonlinear optics,[10,11] metamaterials,[12,13] and transformation optics[14,15] have immensely benefited from the plasmonic properties of metals leading to extraordinary demonstrations of devices such as a perfect lens,[16] optical black holes,[17] and negative refraction[18,19] over the last two decades.

However, most of the aforementioned demonstrations have been performed with noble metals such as gold and silver. Although possessing a high figure of merit (FoM), defined as the ratio of the real part of the permittivity to the imaginary part,[20] they are incompatible with CMOS processes which limits their practicality. Moreover, a limited tunability in the optical properties, an affinity to oxidize under ambient conditions, the formation of deep traps by diffusing into silicon and the low melting points make the incorporation of many traditional metals into microelectronics technology extremely challenging. This has spurred an intense search for robust alternative plasmonic materials.[21,22] In this view, transition metal nitrides have attracted a great deal of interest primarily due to their tunable optical properties, CMOS compatibility and high thermal and chemical stability.[23] Amongst these, titanium nitride (TiN) has been the most widely explored because of its gold like optical properties and refractory nature.[23–30] TiN has already been used in devices demonstrating high efficiency local heating, broadband absorbance in visible and IR and hyperbolic metamaterials with a high photonic density of states.[23,26,31–34] However, sufficiently high quality TiN has generally been grown via sputtering at temperatures exceeding 650°C,[35] which is not acceptable for CMOS processes. To solve this limitation we investigate epitaxial growth of high figure of merit TiN on c-plane sapphire substrates using plasma-enhanced atomic layer deposition (PE-ALD) at temperatures below 500°C. Optical performance of the material is correlated with its structural, electrical and compositional properties.

**Methods:**

Thin films of TiN were grown on c-plane sapphire substrates using PE-ALD (Veeco Fiji G2) equipped with a remote plasma source. Each cycle of ALD consisted of a pulse of Tetrakis(dimethylamido)titanium(IV) (TDMAT) precursor on to the heated sapphire substrate followed by exposure to $N_2$ plasma. The critical growth parameters that were varied were time given to the precursor to chemisorb onto the substrate surface, $t_{chem}$, substrate temperature, $T_s$, and time of plasma exposure, $t_{plasma}$. The exploration of growth parameter space can be broken down to four major series of samples, first of which was targeted at optimizing the chemisorption time, which consisted of films deposited at varying substrate temperatures with two different chemisorption times, $t_{chem}$ = 2s and 0.5s (thickness = 50 nm to 60 nm).[28,37,38] In the second series, the plasma exposure time was optimized by depositing films at different plasma exposure times, $t_{plasma}$ = 10s, 15s, 25s and 35s (thickness = 50 - 60 nm). The third series involved films at thicknesses varying from 30 to 90 nm, grown under the optimum growth conditions established, to investigate thickness dependent optical properties. The final series consisted of films deposited at temperatures varying between $T_s$ = 375°C to 520°C (thickness 90 - 100 nm). Optical characterization was performed using JA – Woollam M-2000 variable angle spectroscopic ellipsometer (VASE) by fitting SE and transmission data simultaneously. Chemical compositions were determined using X-ray photoelectron spectroscopy (XPS) and structural quality of the material was studied using high-resolution ω-2θ and ω-rocking curve X-ray diffraction (XRD) measurements. Resistivity values were determined from Hall effect measurements at 0.5 T. (see supplementary information for further details).

**Results and Discussion:**

Figure 1a shows trend of the peak FoM (varied $T_s$) at $t_{plasma}$ = 10s with two different $t_{chem}$ of 0.5s and 2s, for the first optimization series. A higher FoM is observed for samples deposited with $t_{chem}$ = 0.5s compared to those deposited with $t_{chem}$ = 2s, more so at higher substrate temperatures. This is due to the fact that TDMAT decomposes rapidly at temperatures around 450°C.[39] Thus, the shorter residence time results in reduced premature breakdown of TDMAT. All subsequent films discussed in this work are grown with $t_{chem}$ = 0.5s.

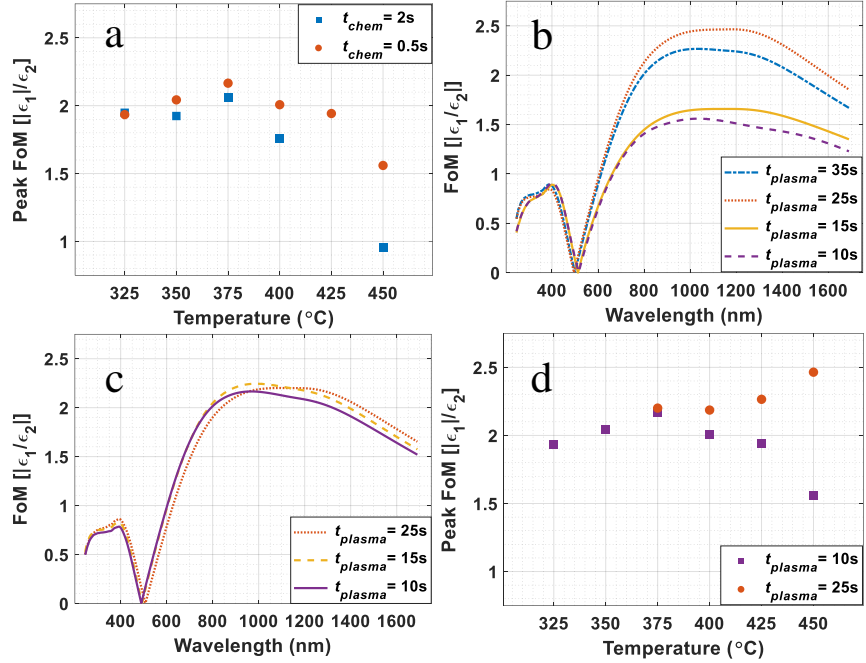

Figure 1: (a) Figure of merit (FoM) of TiN samples grown at varying $T_s$ using $t_{plasma} = 10$s with $t_{chem} = 0.5$s and 2s, (b) FoM of samples grown at $T_s = 450°C$ with varying levels of $t_{plasma}$ using $t_{chem} = 0.5$s (c) FoM of samples grown at $T_s = 375°C$ with varying levels of $t_{plasma}$ using $t_{chem} = 0.5$s. (d) Peak FoM comparison across $T_s$ and $t_{plasma}$ using $t_{chem} = 0.5$s.

Substrate temperature and plasma exposure time play a significant role in the quality of the films.[29] Figure 1b presents a plot of FoM at $T_s = 450°C$ with varying plasma exposure time. It is observed that the FoM increases with an increase in plasma exposure time up to 25s beyond which it decreases. On the other hand, virtually no change in FoM is observed at lower temperatures of $T_s = 375°C$ (Figure 1c). This finding indicates that plasma exposure time has a larger impact on FoM for the films deposited at higher substrate temperatures (>375°C) (Figure 1d). To investigate the genesis of this dependence, we have performed XPS and XRD studies of the films deposited at 450°C with different $t_{plasma}$.

All films regardless of substrate temperature and plasma exposure time exhibited smooth surface morphology (Figure 2a). On the other hand, plasma exposure has been known to reduce impurity concentration in ALD films,[29] which would lead to improvement in optical properties. XPS was conducted on the samples to study the chemical nature of the material. Since TiN tends to form a self-passivated oxynitride layer on the surface,[40] the first 2 nm of the film was sputtered away in the XPS chamber using Ar+ ions. The survey scan indicated the presence of trace amounts of oxygen and carbon along with titanium and nitrogen in the film [Figure S3(a)]. Each of the individual peaks were subsequently decomposed into multiple gaussian peaks and fit to measure the concentration of impurities in the films [Figure S3(b-e)]. Figure 2b shows carbon and oxygen concentrations of films grown at 450°C as a function of $t_{plasma}$. As envisaged, impurity concentrations start to decrease with increasing $t_{plasma}$, until they saturate beyond 35s. This effect is particularly noticeable at temperatures around 450°C because the samples deposited at higher temperatures have a larger concentration of impurities. We attribute this effect to the premature decomposition of the precursor on the substrate surface, wherein the Ti species bond to impurities from the environment. Extended plasma exposure facilitates complete conversion of Ti into stoichiometric TiN. Samples grown at $T_s$ = 375°C with $t_{plasma}$ = 10s had less than 2±0.5% carbon and oxygen impurities as compared to 4% carbon and 3% oxygen impurities in the films grown at $T_s$ = 450°C. With all the samples being nearly stoichiometric [Figure S3(f)], the decrease in impurity content led to enhanced optical performance up to $t_{plasma}$ = 25s; however, the reduction in impurity content fails to explain the drop in the FoM value observed for the film grown with $t_{plasma}$ = 35s at 450°C. This drop has its origins in reduced structural quality of the films deposited with long plasma exposures, as evidenced from XRD measurements.

The XRD ω-2θ patterns reveal the presence of only (111) and (222) reflections from TiN at 36.7° and 73.2°, respectively, which together with the ϕ-scans (not shown) illustrates the epitaxial growth of TiN on c-plane sapphire (Figure 2c). However, the decrease in intensity of the ω-2θ peaks (Figure 2c) along with broadening of XRD ω-rocking curves (full width at half maximum values are 0.17°, 0.35°, 1.9°, and 2.3° for plasma exposures of 10, 15, 25, and 35 s, respectively) of the corresponding samples indicates the increase in mosaicity of the films with an increasing plasma exposure time. This in turn is related to smaller and more misaligned grains along with an increase in point defect density, which causes an increased carrier collision rate and optical losses. The increase in defect density is due to films being marginally nitrogen rich at higher plasma exposure times [Figure S3(f)]. The excess nitrogen and the increase in mosaicity are believed to be caused by plasma damage (surface bombardment by high energy plasma species) which increases with longer exposure times, a trend which has been noted in other plasma enhanced deposition methods, such as CVD and MBE.[41,42] Therefore, a trade-off between reduced impurity concentration and increased defect density with plasma exposure time gives rise to the best optical properties of the TiN films deposited with $t_{plasma}$ = 25 s.

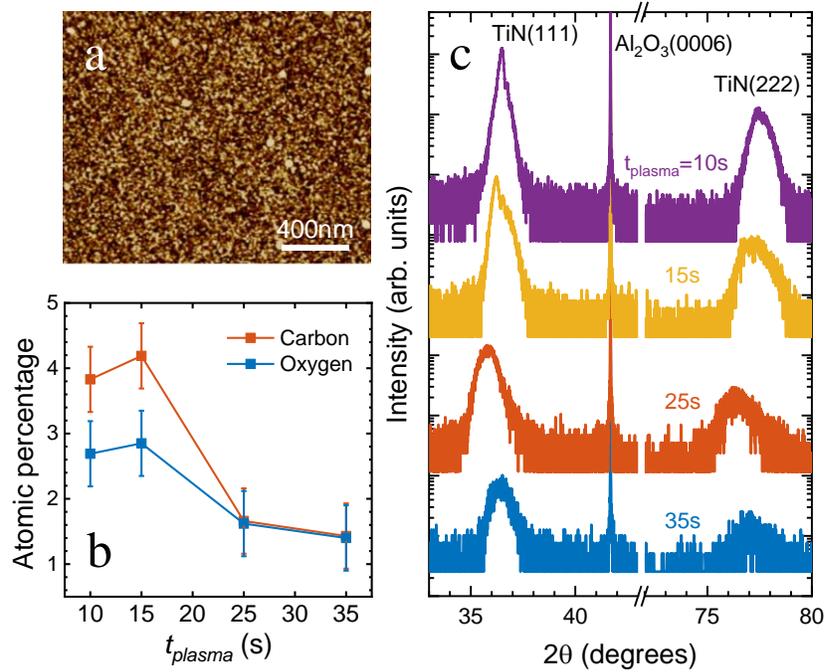

Figure 2: (a) Atomic force microscopy image scan of the film deposited at 450°C with $t_{plasma}$ = 25 s indicating smooth surface morphology (RMS value is <1 nm, Δz = 2 nm) (b) Atomic percentage of carbon and oxygen in samples grown at $T_s$ = 450°C with varying $t_{plasma}$ (c) XRD ω-2θ scans of the corresponding samples.

Finally, as permittivity of the optically thin films is affected by the substrate and the near-interface layer of reduced quality (substrate-TiN interface), we performed a thickness dependent study of FoM. Figure 3a shows a rise in peak FoM from 2.2 to 2.8 with film thickness increasing from 30 nm to 80 nm, beyond which it saturates. The observed dependence is due to contribution from the lower quality layer near the substrate/TiN interface. Beyond a few skin depths (δ ~ 15 nm at the frequency corresponding to the peak FoM), this contribution becomes negligible, as indicated by FoM saturation at thickness >80 nm (Figure 3a). Figure 3b shows the peak FoM of optically thick films (~100nm), as a function of substrate temperature. As seen from the figure, $T_s$ = 450°C is the optimal deposition temperature, which also produces the lowest film resistivity of 31 μΩ - cm. A sharp drop in FoM accompanied by a rapid increase in growth rate, from 1.3 A/cycle at 450°C to 4.2 A/cycle at 500°C is observed, thus indicating CVD-like growth at high temperatures. Figure 3c and 3d show the real and imaginary permittivity along with the FoM of the optimized film.

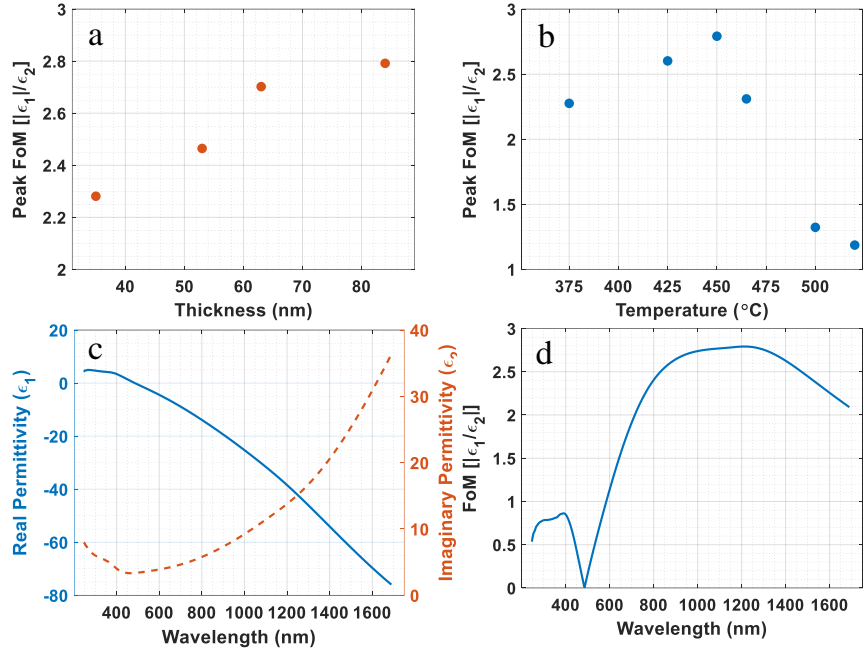

Figure 3: (a) Peak FoM for samples grown under conditions of $t_{chem} = 0.5s$, $t_{plasma} = 25s$ and $T_s = 450°C$, (b) Peak FoM for optically thick samples (5 to 6 skin depth thickness), $T_s = 450°C$ is the optimal substrate temperature, (c) Real and imaginary permittivity for the film grown under optimized conditions, (d) FoM of the optimized film, peaking at 2.8.

To demonstrate the high plasmonic quality TiN film grown via PE-ALD, we have explored extraordinary transmission (EOT) through sub-wavelength apertures, which although fundamentally possible in all metallic films, can be observed in only experiments on high q-factor plasmonic films[43–45]. Briefly, EOT is a drastic enhancement of the transmission through a metallic film with an etched periodic lattice of sub-wavelength apertures that is due to the excitation of localized surface plasmons around the holes, which tunnel through the sub-wavelength aperture.[45] In this work, structures with a diameter of 300 nm and periodicity of 500 nm were simulated using a finite element method solver (COMSOL) to obtain a broad peak around 1150 nm, which falls in the near IR transparency window of biological tissues. The electric field profile at 1150 nm is shown in subset of Figure 4a. Subsequently, the structures were fabricated using a 45 nm thick film grown under optimized conditions ($T_s = 450°C$, $t_{plasma} = 25s$). 45 nm thick films were used to improve the overall transmission of the EOT peak, as compared to the optimized 80 nm thick films. The film was spin coated with 300 nm thick ZEP-520A followed by baking at 180°C for 180s. Then the structures were written using electron beam lithography with a 75 μC/cm$^2$ dose over a 90 μm by 90 μm write area. After development in Xyelene for 60s, 300 nm apertures were etched using an inductively coupled plasma reactive ion etching system (ICP-RIE) in a $Cl_2$/Ar gas chemistry using ZEP-520A as the mask. The SEM image of the metasurface is shown in Figure 4a. Figure 4b shows the simulated and the experimental transmission spectra. The peak in transmission around 1150 nm through subwavelength 300 nm diameter holes is attributed to the excitation of localized surface plasmons (LSPR) on the film, which tunnel through and couple out to free space on the other side. These LSPR induced EOT peaks were nonexistent in low quality ALD films (peak FoM < 1.5). The resulting spectrum is broader than the simulated one because of the roughness on sidewalls and other fabrication non-idealities.

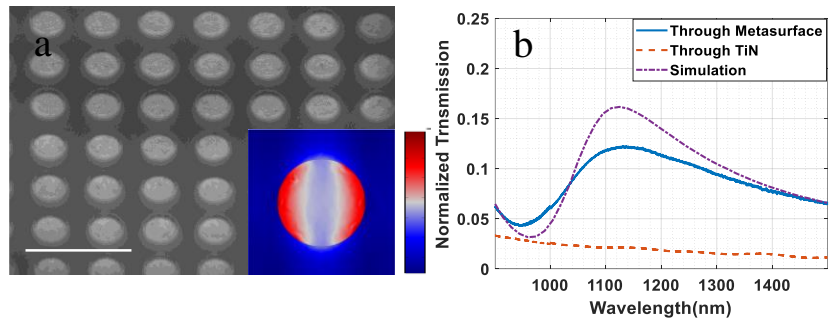

Figure 4: (a) SEM image of the fabricated sub-wavelength aperture with a scale bar of 1µm, inset shows the electric field profile through the subwavelength aperture at 1150 nm, the scale for which varies from 0 to $1.5 \times 10^8$ V/m. (b) Simulated and measured transmission spectra through the sub-wavelength periodic grating along with the transmission spectra through bare titanium nitride film.

**Conclusion:**

To summarize, TiN films were epitaxially grown on c-plane sapphire substrates using PE-ALD. Temperatures higher than those typically used in ALD processes but within the realm of CMOS technology were explored, which led to significant improvement in the FoM. Plasma exposure time, substrate temperature, and chemisorption time were optimized to obtain the material with the best optical properties. The highest FoM of 2.8 was observed in samples grown at 450°C with a 0.5s chemisorption time followed by 25s plasma exposure at film thickness greater than 80nm, demonstrating high plasmonic quality, CMOS compatible robust titanium nitride films. Beyond 25s, a degradation of optical properties was observed because of the increased defect density in the films. Although an increase in plasma exposure time up to 25s resulted in enhancement of electrical and optical properties and improved stoichiometry of the TiN films, it results in broadening of XRD rocking curves likely caused by increased mosaicity and defect density. We predict that switching to alternative precursor(s) which are stable at higher temperatures such as TDEAT could result in films with better optical quality. Finally, EOT was demonstrated through sub-wavelength apertures of 300 nm diameter and 500 nm periodicity in a biocompatible, tunable and thermally robust film grown via a low-cost, scalable and conformal approach, thus bringing TiN a step closer to practical applications such as in heat assisted magnetic recording devices (HAMR).

**Supplementary Material**

See supplementary material for ALD parameters, ellipsometer model, XPS and resistivity data.

**Acknowledgments**

The work was supported by a grant from the Virginia Microelectronic Consortium (VMEC).

# Supporting Information For

# Plasmonic Titanium Nitride via Atomic Layer Deposition: A Low-temperature Route

Dhruv Fomra[1], Ray Secondo[1], Kai Ding[1], Vitaliy Avrutin[1], Natalia Izyumskaya[1], Ümit Özgür[1], Nathaniel Kinsey[1]

Department of Electrical and Computer Engineering, Virginia Commonwealth University, Richmond, Virginia – 23220

**S1. ALD Parameters**

Atomic layer deposition is a self-limiting vapor phase technique (see Figure S1 (a)), which typically offers highly controllable low growth rates of 0.1 to 0.5 nm per cycle in case of titanium nitride. A wide variety of precursors, including Tetrakis(dimethylamino)titanium (TDMAT), Tetrakis(diethylamino)titanium (TDEAT), and titanium tetrachloride ($TiCl_4$) have been used for ALD growth of TiN, with TDMAT being the most used one. It has moderate decomposition temperatures and a large ALD-window temperature range, enabling large tunability of optical properties. The ALD window is essentially the temperature range across which one observes self-limiting growth mechanism. At temperatures below the ALD window, the precursor could condense on the surface of the substrate, leading to ineffective purging and resulting in an increased growth rate. On the other hand, it could also happen that due to low temperature, the chemisorption step remains incomplete, which would result in a decreased growth rate. On the other end of the ALD window, at higher temperatures, decomposition of the precursor could lead to a either a CVD type growth where the growth rate can be increased or enhanced desorption of metalorganic precursor which may decrease the growth rate (Figure S1 (b)). The ALD parameters used in our recipes

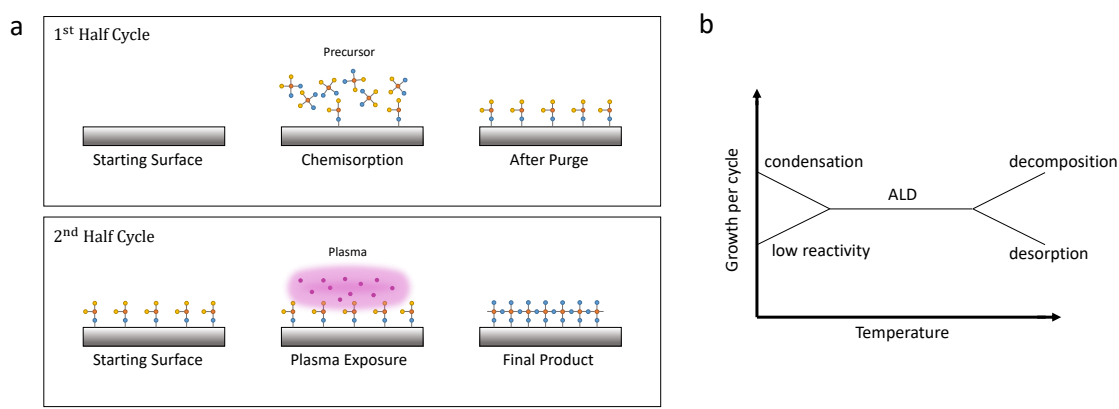

Figure S1: (a) Schematic of PE-ALD cycle, showing the metalorganic precursor pulse chemisorbing onto the surface of the heated film, followed by plasma exposure which provides the activation energy for the formation of titanium nitride (b) Schematic showing the ALD temperature window and transition to CVD regimes at higher temperatures

involved temperatures of the precursor source, delivery line, and reactor walls of 75°C, 150°C, and 300°C, respectively, while the substrate temperature ranged from 350°C to 500°C. The plasma exposure time was varied between 10s and 35s. At this point, it is important to note that the process was carried out at substrate temperatures bordering the ALD window on high-temperature side, near the transition to a CVD growth mode. Higher temperatures were explored to obtain better structural quality of films.

**S2. Ellipsometer Drude-Lorentz fits**

Optical characterization of the films was performed using JA –Woollam M-2000 variable angle spectroscopic ellipsometer (VASE) by fitting SE and transmission data simultaneously. The thicknesses of a set of samples were also verified independently via etching experiments[1]. The data from the ellipsometer was fit using a B-Spline model enforcing causality via Kramers-Kronig relation and was later fit to a Drude-Lorentz model to extract physicality from the fit (Table S1). B-spline was used to model the optical properties of TiN, which gives complete flexibility to the optical constants unlike oscillator-based models. To maintain Kramers-Kronig (KK) consistency, Complete Ease (software) fits the psi and delta values to the imaginary component of the permittivity, and subsequently calculates the real part of it by KK relation. An alternative way is to use the more limited but physical Drude-Lorentz model. Based on the interband transitions seen in the band structure of titanium nitride[2] at approximately 3 and 6 eV, the data from the ellipsometer is fit to the Drude model with the parameters listed in Table S1.

Table S1: Parameters used for Drude-Lorentz modelling

| | Drude | | | Lorentz | | |
|---|---|---|---|---|---|---|
| | Carrier Concentration (cm$^{-3}$) | Gamma (fs) | Effective Mass | Amplitude | Broadening | Energy (eV) |
| **Drude** | 3.55×10$^{22}$ | 3.01 | 0.863 | - | - | - |
| **Lorentz 1** | - | - | - | 8.944 | 4.0548 | 6.151 |
| **Lorentz 2** | - | - | - | 1.291 | 0.9364 | 3.598 |

## S3. Compositional Analysis – X-ray Photoelectron Spectroscopy

X-ray photoelectron spectroscopy was conducted on the sample grown at 450°C with 25 s plasma exposure, to study the chemical composition of the material. Since TiN tends to form a self-passivated oxynitride layer on the surface, the top 2 nm of the film was etched away sputtered with Ar+ ion beam. Figure S3a indicates the presence of trace amounts of oxygen and carbon in the TiN film. Each of the individual peaks were subsequently deconvoluted into multiple Gaussian peaks. The $C_{1s}$ spectrum is unfold into 2 peaks, indicating the presence of titanium carbide and aliphatic carbon, with peaks at 282.1 eV and 284.4 eV, respectively (Figure S3d). The $O_{1s}$ spectrum corresponds to a single peak at 532. 2 eV of Ti-O-N bond, indicating the presence of oxynitride (Figure S3e). The presence of Ti-O-N is further confirmed by the nitrogen spectrum that, shows a broad peak at 398.5 eV accompanied by a dominant peak at 397.4 eV due to Ti-N bonding[3–5] (Figure S3c). Finally, the titanium spectrum supports the presence of Ti-N and Ti-O-N bonds with peaks at 455.1 and 456.2 eV, respectively, although the peak corresponding to Ti-C, which usually observed at 454.9 eV, is probably obscured by the low-energy wing of strong 455.1-nm line (Figure S3b).

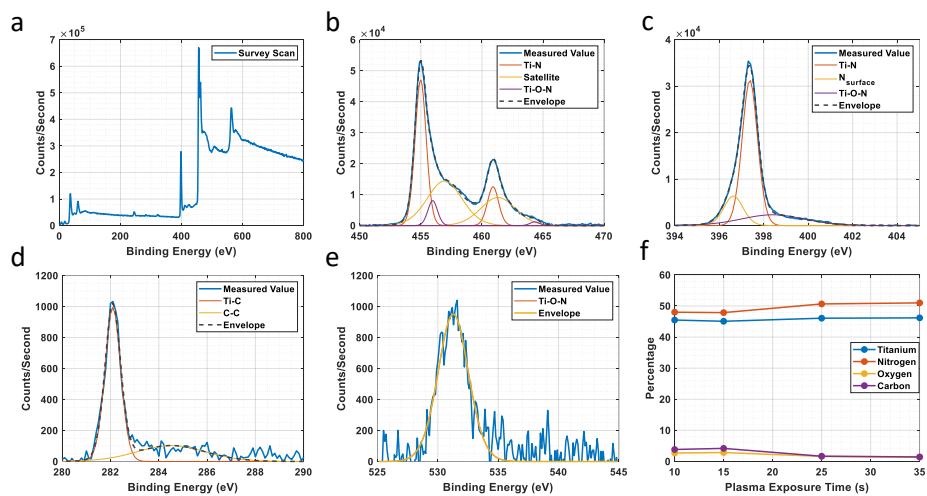

Figure S3: (a) XPS survey spectrum of the TiN films grown at 450°C and 25s plasma exposure on c-plane sapphire, (b) Ti2p, (c) N1s, (d) C1s and (e) O1s peaks fitted to determine the chemical nature of bonds of the elements present in the film, (f) Chemical composition of the films, showing nearly stoichiometric films.

## S4. Resistivity

Figure S4 shows resistivity values determined from the Hall effect measurements as a function of substrate temperature for the 50 nm TiN films deposited with 10 and 25s plasma exposure times. The resistivity reduces by 30% for the film deposited with 25s plasma exposure at 450°C as compared to that in the layers grown with 10s plasma exposure, although plasma exposure time has virtually no effect on the resistivity of films deposited at ≤400°C. This is because of the decomposition of the precursor at elevated temperatures, which increases the concentration of impurities in the samples. In this sense, for $T_s>400°C$, $t_{plasma}$ must be increased to prevent the Ti species bonding with residual impurities in the environment, which thereby improves the quality of the films. Also, increasing the thickness to 85 nm reduced the resistivity further to 31 µΩ-cm.

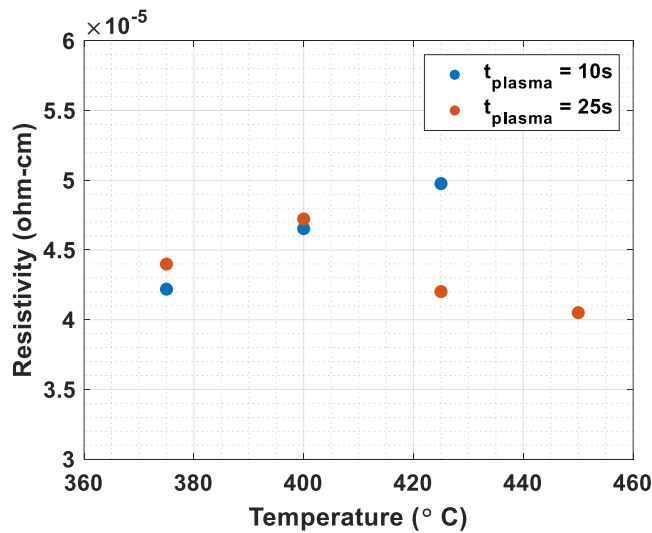

Figure S4: (a) Resistivity vs temperature for samples grown with 10s and 25s plasma exposure times. Resistivities as low as 41 µΩ − cm have been obtained for the thin 50 nm to 60 nm films.